# Magnetic behavior of nanocrystalline ErCo$_2$

Sitikantha D Das, Niharika Mohapatra, Kartik K Iyer, R.D. Bapat, and E.V. Sampathkumaran[*]
Tata Institute of Fundamental Research, Homi Bhabha Road, Colaba, Mumbai 400005, India

We have investigated the magnetic behavior of the nanocrystalline form of a well-known Laves phase compound, ErCo$_2$ - the bulk form of which has been known to undergo an interesting first-order ferrimagnetic ordering near 32 K – synthesized by high-energy ball-milling. It is found that, in these nanocrystallites, Co exhibits ferromagnetic order at room temperature as inferred from the magnetization data. However, the magnetic transition temperature for Er sublattice remains essentially unaffected as though the (Er)4f-Co(3d) coupling is weak on Er magnetism. The net magnetic moment as measured at high fields, say at 120 kOe, is significantly reduced with respect to that for the bulk in the ferrimagnetically ordered state and possible reasons are outlined. We have also compared the magnetocaloric behavior for the bulk and the nano particles.

## I. Introduction

The Laves phase compounds, $RCo_2$ (R= Rare-earths), attracted considerable attention in the literature during last few decades, as this family served as a model system for itinerant electron metamagnetism (IEM) [1]. For instance, $YCo_2$ and $LuCo_2$, the exchange-enhanced Pauli-paramagnets, have been found to undergo IEM at about 700 kOe [2, 3]. With respect to those members in which R carries a localized-moment, quite interestingly, the magnetic transition appears to be first order for R= Dy, Ho, and Er, whereas for many other members of this series, it is second order. It is generally accepted that the first-order nature of the transition for the former members is due to the onset of IEM at the Co site induced by the molecular field arising from the magnetic ordering of R sub-lattice. This finding led to intense investigations [4], particularly with respect to potential magnetic refrigeration applications [5].

While all the reports on these heavy rare-earth members is on bulk form (single crystals and poly crystals), very little work has been carried out on nanocrystalline form, as it is not usually easy to stabilize the nanoparticles of rare-earth intermetallics without shell protection. In this respect, there is a recent claim on the synthesis of oxidation-resistant $DyCo_2$ nano particles by arc-discharge process [6] with a promising low-temperature refrigeration applications. In this article, we present the results of our magnetic investigation on the nanoparticles of $ErCo_2$ synthesized by high-energy ball-milling [7], the reason being that this compound in particular attracted considerable attention in the recent literature [8-21]. This compound has been found to order ferrimagnetically at ($T_C$=) 32 K due to antiparallel aligment of Er and Co moments (with a moment on Co being about $1\mu_B$) [8] with extreme sensitivity of $T_C$ to small amounts of metallic impurities [9, 10]. This compound is well-studied for its magnetocaloric effect (MCE) [11-13]. The so-called 'inverse IEM' phenomenon has been noted for some degree of chemical doping [20].

## II. Experimental details

The bulk sample was prepared by arc melting stoichiometric amounts of high-purity (>99.9%) constituent elements in an atmosphere of argon. The molten ingot was annealed at 900 C for 60 h in an evacuated sealed quartz tube. The ingot was then subjected to high-energy ball-milling (Fritsch pulverisette-7 premium line) for 2 ½ h employing zirconia vials with balls of 5mm diameter (balls-to-material mass ratio: 5) with an operating speed of 500 rpm in an atmosphere of toluene. The specimens were characterized by x-ray diffraction (XRD), scanning electron microscope (SEM, JEOL, JSM 840A), transmission electron microscope (TEM, Technai 200 kV) and energy-dispersive x-ray analysis. A commercial magnetometer (Quantum Design) was employed to measure magnetization.

## III. Results and discussion

The XRD patterns are shown in figure 1 for both the ingot and milled specimens. The patterns are shifted along y-axis for the sake of clarity. Otherwise the background intensity and shape of the background are found to be essentially the same for both the specimens, thereby indicating that the milling does not result in amorphous formation. The sharp XRD lines in addition endorse that the milled samples are crystalline. We did not find any noticeable change in the lattice constants (as determined from XRD) due to ball-milling. As well-known in the field of metallurgy, the milling reduces the intensity of the XRD lines. The XRD lines are broadened as demonstrated in an inset of figure 1 and the average particle-size, if estimated from the width of the most intense line, turns out to be about 30 nm. The nano-sized nature of the specimen was ascertained from TEM pictures (see figure 1). However, the TEM picture suggests that the particle size is smaller than that estimated from XRD. We attribute this discrepancy to the fact that the strain-contribution to broadening of XRD lines could not be separated due to non-linear Williamson-Hall plots (not shown here). The electron diffraction pattern inserted in figure 1 confirms that the particles belong to $ErCo_2$ phase. We have also carried out Reitveld fitting of the XRD pattern for the milled specimen and the difference between the experimental and fitted pattern is shown in figure 1 to endorse proper phase formation. There is no evidence in the XRD for the formation of any other phase due to ball-milling. From the backscattered image of SEM, we further confirmed the absence of any other phase in the nanospecimen. In addition, the stoichiometry of the nanospecimen was established to correspond to $ErCo_2$ from energy dispersive x-ray analysis performed with this SEM instrument.

The results of magnetization measurements as a function of temperature ($T$) and magnetic-field ($H$) for the zero-field-cooled (ZFC from 300 K) and field-cooled (FC) conditions of the milled specimen are shown in figures 2 and 3 respectively. The data obtained on the parent ingot is also included for comparison. In the ingot employed to prepare the fine particle, we see the features attributable to the onset of magnetic ordering near 36 K as inferred, say, from the peak temperature in the $M(T)$ obtained in a field of 100 Oe. It is straightforward to see the changes (see figure 2) those have occurred in the shapes of $M(T)$ curves after milling. Most notably, the sharpness of the features seen at $T_C$ for the bulk specimen is absent for the nano particles; see, for instance, the features in the ZFC curve ($H$= 100 Oe). This is attributable to the broadening of the magnetic transition due to defects introduced by ball-milling. In addition to above features near magnetic transition temperature, there is a bifurcation of ZFC-FC curves extending beyond 36 K in the nano specimen, which is indicative of another magnetic anomaly at higher temperatures. This aspect is addressed in the next paragraph on the basis of $M(H)$ anomaly.

For the bulk specimen, the $M(H)$ plots (figure 3) are hysteretic below 20 kOe in the magnetically ordered state and the value of $M$ tends towards saturation beyond 30 kOe with the saturation moment (say, about 7.2 $\mu_B$ per formula unit at 1.8 K, see also Ref. 13) being less than that expected for trivalent Er ions due to ferrimagnetism. As expected, in the paramagnetic state, say, at 50 K, the plot of $M(H)$ is linear. On the other hand, in the nanoform, hysteretic nature of the curve persists even near 50 kOe, for instance, at 1.8 and 30 K (see figure 3) without any evidence for saturation; in addition, high-field magnetic moment is significantly reduced. On the basis of these $M(H)$ properties, we infer that ferrimagnetism is retained in the nano form. If one performs a linear extrapolation of the high-field data to zero-field at 1.8 K, assuming that the moment on Er is unchanged, we arrive at a value of at least 2 $\mu_B$ on each Co ion for the nanoform. Thus, if one solely attributes the observed reduction of high-field magnetic moment to Co, then there is an enhancement in the magnetic moment on Co following Er sub-lattice ordering. However, it is also possible that a significant surface spin disorder due to lack of symmetry and reduced coordination at the surface is also responsible for the observed behavior, as known, for instance, for the nanoform of γ-$Fe_2O_3$ [22]. The finding we stress for the nanoform is that, well above 36 K, the $M(H)$ plots show a dramatic increase for initial applications of magnetic field with a linearity at higher fields, as though there is a ferromagnetic component superimposed over a paramagnetic component. The plots are hysteretic at low fields as shown for 130 K and 300 K in the inset of figure 3. These findings reveal that Co in the nanoparticle exhibits ferromagnetic character at room temperature with Er remaining paramagnetic. The value of the magnetic moment on Co turns out to be ~0.12 $\mu_B$/Co, at least above 32 K. We believe that when Er is paramagnetic, it is possible that Co spins undergo 'canted' ferromagnetic alignment, as magnetic moment value is lower than that observed [7] in $YCo_2$. In support of a complex magnetic ordering of Co, the plot of $M/H$ exhibits an increase above 200 K as shown in the inset of figure 2. It is of interest to carry out neutron diffraction studies to understand this issue better. In any case, it is clear that there is a magnetic ordering of Co in the nano particles of $ErCo_2$ above room temperature. It is to be stressed that, despite high magnetic ordering temperature of Co, the $T_C$ for the Er sublattice is not noticeably influenced.

We find dramatic modifications of Arrott plots in the temperature range where Er sub-lattice undergoes magnetic ordering (see figure 4). For the ingot, these plots have been known to be complex with negative slopes and inflexions due to metamagnetic transitions the exact nature of which appears to be sample dependent [see, for instance, Refs. 13, 22]. In the nano specimen, though $M^2$ increases monotonically with $H/M$, the plot is still complex which is attributable to a possible interference from Co sub-lattice ordering at higher temperatures.

We compare MCE properties in figure 5 for a typical change of magnetic field. For this purpose, we in fact obtained $M(H)$ curves at close intervals of temperatures and derived entropy change, $\Delta S$, for a given variation of $H$ (from zero field), on the basis of well-known Maxwell's relation (see, for instance, Ref. 5). The results are shown in the figure for selected fields. The results obtained for the ingot are in good agreement with those known in the literature [11-13, 21]. It is distinctly clear from figure 5 that, in the case of nanospecimen, though $\Delta S$ values at the peak are relatively reduced at $T_C$, the curve is broad extending over a wider temperature range. The magnetic refrigeration capacity [for the definition we use, see Ref. 23], turns out to be about half of that seen for the bulk form [about 80 J/mol].

### IV. Conclusion

The compound, $ErCo_2$, attracted attention in the literature from the angle of potential magnetic refrigeration applications as well as of interesting magnetic anomalies. We have synthesized the nanocrystals of this material, to our knowledge, for the first time and studied its magnetic behavior. The Co sublattice is found to be magnetically ordered at room temperature in these fine particles unlike in bulk. It is notable that the transition temperature for the Er sub-lattice is not influenced at all by the high-temperature magnetism of Co sub-lattice, as though Er-Co magnetic coupling is weak in the nanoform. There is a significant reduction in the high-field magnetic moment compared to that of bulk when Er sub-lattice orders at low temperatures. Though the magnetic refrigeration capacity becomes half of the bulk material in the temperature range of interest, the entropy change is spread over a rather larger temperature range when compared to that in bulk form.

This family of binary compounds are characterized by sharp density of 3d states of Co in the vicinity of Fermi level and, in the case of pseudo-binary compounds based on $YCo_2$, even defects have been proposed to have a profound effect on this feature and hence on magnetism [24]. It is not clear whether similar effect is responsible for the magnetism of Co in the ball-milled $ErCo_2$ as well. Thus, it appears that this family could be a good example for probing an interplay between the defects, electronic structure and strong electron correlations.

[*]E-mail: sampath@mailhost.tifr.res.in
1. E.P. Wohlfarth and P. Rhodes, Philos. Mag. **7**, 1817 (1962).
2. T. Goto, K. Fukamichi, T. Sakakibara, and H. Komatsu, Solid State Commun. 72, 945 (1989).
3. H. Yamada, T. Tohyama and M. Shimizu J. Magn. Magn. Mater. **70,** 44 (1987); T. Yokoyama et al., J. Phys.: Condens. Matter. **13**, 9281 (2001).
4. S. Khmelevskyi and P.Mohn, J. Phys.: Condens. Matter **12**, 9453 (2000); N.H. Duc and T. Goto In: K.A. Gschneidner, Jr. and L. Eyring, Editors, *Handbook on the Physics and Chemistry of Rare Earths* **Vol. 28**, North-Holland, Amsterdam (1999), p. 177 and references therein.
5. See, for a review, K.A. Gschneidner Jr, V.K. Pecharsky, and A.O. Tsokol, Rep. Prog. Phys. **68**, 1479 (2005).
6. S. Ma, W.B. Cui, D. Li, N.K. Sun, D.Y. Geng, X. Jiang, and Z.D. Zhang, App. Phys. Lett. **92,** 173113 (2008).
7. Through this procedure, in the past, we were able synthesize stable form of nano particles of $YCo_2$, with ferromagnetism at room temperature. See, S. Narayana Jammalamadaka, E.V. Sampathkumaran, V. Satya Narayana Murthy, and G. Markandeyulu, App. Phys. Lett. **92,** 192506 (2008).
8. M. Moon, W.C. Koehler, and J. Farrell, J. Appl. Phys. **36,** 978 (1965).
9. X.B. Liu and Z. Altounian, J. Appl. Phys. **103,** 07B304 (2008).
10. M. Guillot and Y. Öner, J. Appl. Phys. **103**, 07E137 (2008).
11. T.D. Cuong, L. Havela, V. Sechovsky, A.V. Andreev, Z. Amold, J. Kamarad, and N.H. Duc, J. Appl. Phys. **81,** 4221 (1997).
12. H. Wada, S. Tomekawa, and M. Shiga, Cryogenics **39,** 915 (1999).
13. A. Gigurre, M. Foldeaki, W. Schnelle, and E. Gmelin, J. Phys.: Condens. Matter **11,** 6969 (1999).
14. J. Herrero-Albillos, D. Paudyal, F. Bartolome, L.M. Garcia, V.K. Pecharsky, K.A. Gschneidner, Jr., A.T. Young, N. Jaouen, and A. Rogalev, J. Appl. Phys. **103,** 07E146 (2008).
15. N. Ishimatsu, S. Miyamoto, H. Maruyama, J. Chaboy, M.A. Laguna-Marco, and N. Kawamura, Phys. Rev. B 75, 180402(R) 2007.
16. O. Syshchenko, V. Sechovsky, M. Divis, T. Fujita, R. Hauser, and H. Fujii, J. Appl. Phys. **89**, 7323 (2001).
17. O. Syschenko, T. Fujita, V. Sechovsky, M. Divis, and H. Fujii, Phys. Rev. B **63**, 054433 (2001).
18. R. Hauser, E. Bauer, and E. Gratz, Phys. Rev. B **57**, 2904 (1998).

19. A. Podlesnyak, Th.Strässle, J. Schefer, A. Furrer, A. Mirmelstein, A. Pirogov, P. Markin, and N. Baranov, Phys. Rev. B **66,** 012409 (2002).
20. R. Hauser, E. Bauer, E. Gratz, H. Mueller, M. Rotter, H.Michor, G. Hilscher, A.S. Markosyan, K. Kamishima, and T. Goto, Phys. Rev. B **61**, 1198 (2000).
21. Z. Jun-Ding, Shen Bao-Gen, and S. Ji-Rong, Chinese Physics **16,** 1817 (2007).
22. T.N. Shendruk, R.D. Desautels, B.W. Southern, and J. van Lierop, Nanotechnology, **18,** 455704 (2007); E. Tronc, D. Fiorani, M. Nogues, A.M. Testa, F. Lucari, F.D'Orazio, J.M. Greneche, W. Wernsdorfer, N. Galvez, C. Chaneac, D. Mailly, and J.P. Jolivet, J. Magn. Magn. Mater. **262,** 6 (2003).
23. N. Mohapatra, Kartik K Iyer and E.V. Sampathkumaran, Eur. Phys. J. B **63**, 451 (2008).
24. A.T. Burkov, E. Bauer, E. Gratz, R. Resel, T. Nakama, and K. Yagasaki, Phys. Rev. B **78**, 035101 (2008).

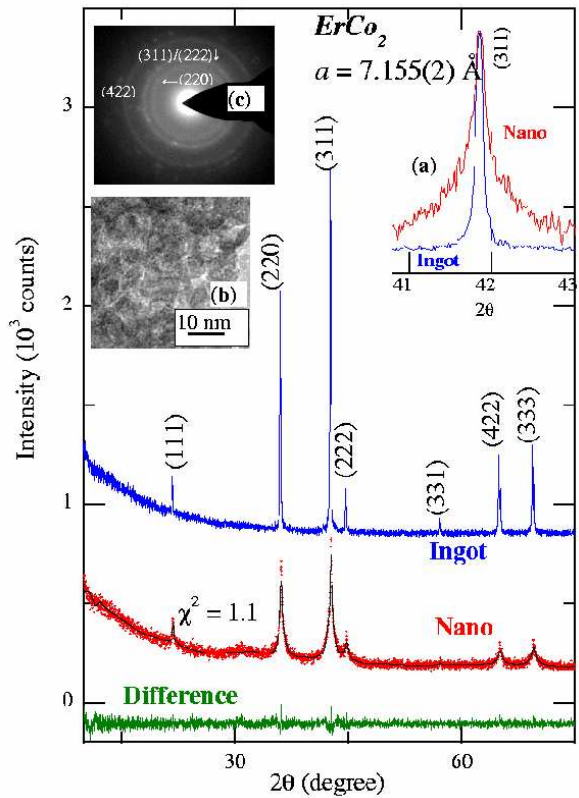

Figure 1:
(color online) X-ray diffraction patterns of the bulk and nano crystals of $ErCo_2$. The fit obtained by Reitveld analysis in the case of nanospecimen is shown by continuous line and the difference between fit and experimental data points are also shown. In the inset **(a),** the shapes of most intense lines are compared after normalizing to respective peak heights. In the insets **(b)** and **(c),** TEM image and electron diffraction pattern are shown.

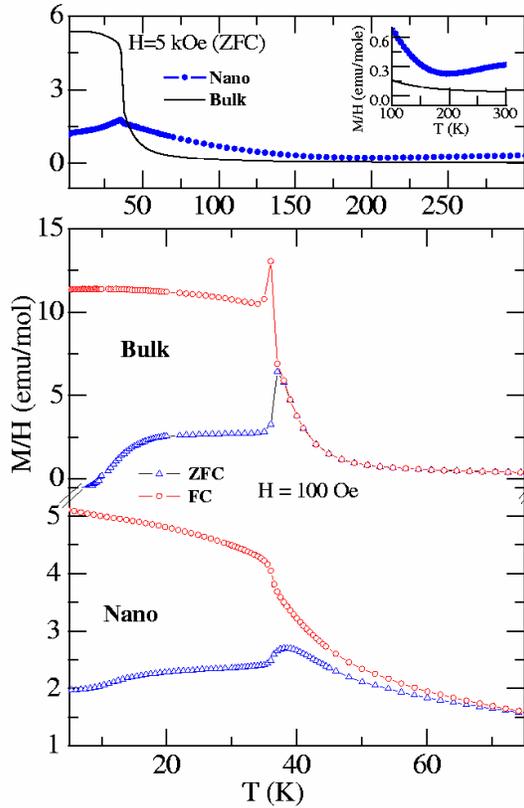

Figure 2:
(color online) Magnetization (*M*) divided by magnetic field (*H*) as a function of temperature (*T*) for the bulk and nanocrystals of $ErCo_2$ for two values of externally applied magnetic fields. In the case of *H*= 100 Oe, the curves for the zero-field-cooled and field-cooled conditions of the nanoparticles are shown. The lines through the data points serve as guides to the eyes. Inset highlights increasing *M/H* behavior above 200 K for the nanocrystals (shown along with the data for the bulk).

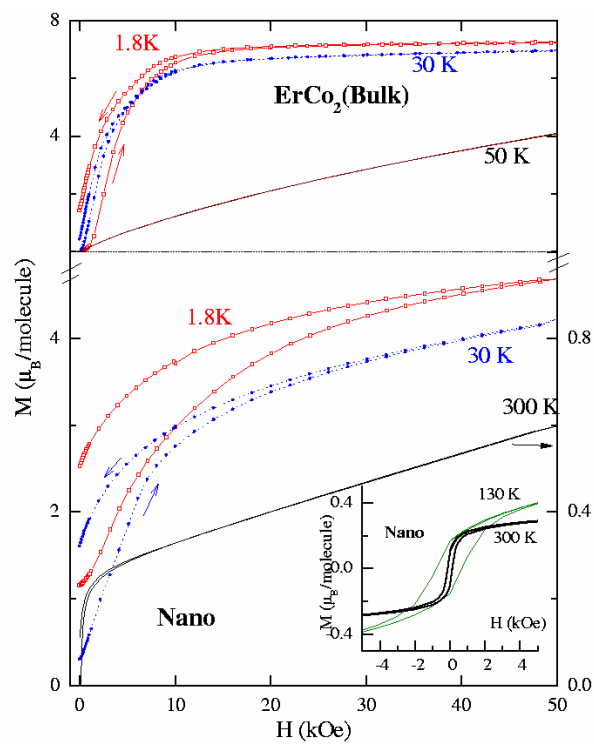

Figure 3:
(color online) Isothermal remnant behavior at selected temperatures for the bulk and nanocrystals of $ErCo_2$. The lines through the data points serve as guides to the eyes. In the inset, low-field hysteresis loops at 130 and 300 K for the nano crystals are plotted.

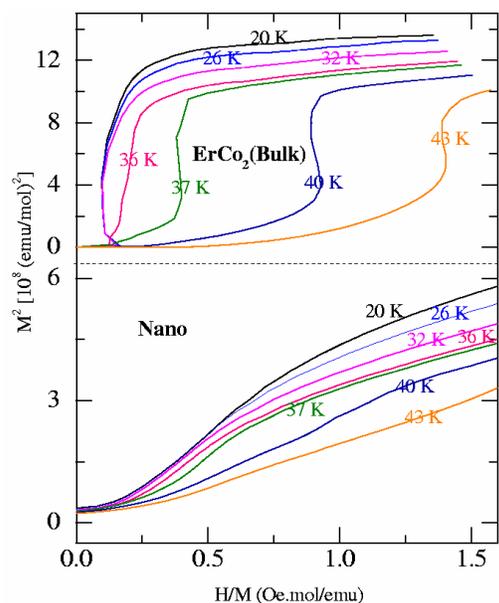

Figure 4:
(color online) The Arrott plots for bulk and nanocrystals of $ErCo_2$. For the sake of clarity, we show only the lines through the data points.

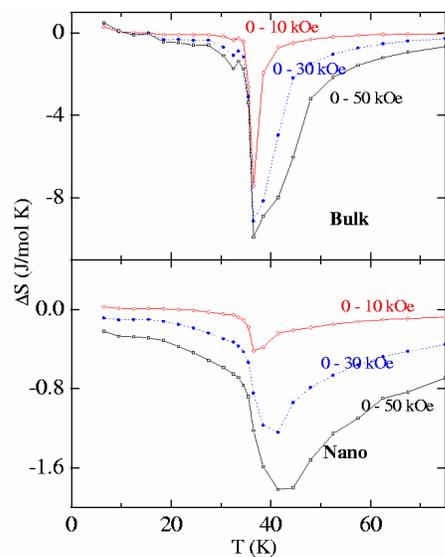

Figure 5:
(color online) The temperature dependence of entropy change for different variations of magnetic fields for the bulk and nanocrystals of $ErCo_2$. The lines through the points serve as guides to the eyes.